**Artificial Intelligence in Clinical Health Care Applications: Viewpoint**

Michael van Hartskamp, Sergio Consoli, Wim Verhaegh, Milan Petković, Anja van de Stolpe

Philips Research Eindhoven, High Tech Campus, 5656 AE Eindhoven, The Netherlands

Corresponding author: Anja van de Stolpe, anja.van.de.stolpe@philips.com

**Abstract**

The idea of Artificial Intelligence (AI) has a long history. It turned out, however, that reaching intelligence at human levels is more complicated than originally anticipated. Currently we are experiencing a renewed interest in AI, fueled by an enormous increase in computing power and an even larger increase in data, in combination with improved AI technologies like deep learning. Healthcare is considered the next domain to be revolutionized by Artificial Intelligence. While AI approaches are excellently suited to develop certain algorithms, for biomedical applications there are specific challenges. We propose recommendations to improve AI projects in the biomedical space and especially clinical healthcare.

**Introduction**

The idea of Artificial Intelligence (AI) has a long history. Since the 1950's there have been several revolutionary promises of AI replacing human work within a few decades. It turned out, however, that reaching intelligence at human levels was more complicated, and this led to several 'AI winters' where interest in AI disappeared[1]. Currently we are experiencing a renewed interest in AI, fueled by an enormous increase in computing power and an even larger increase in data generation. In combination with improved algorithms that allow training of deep neural networks, several high-tech companies have reached successes in performing tasks that are close to human or even beyond human performance: playing games like Chess and Go, image recognition and computer vision, natural language processing, machine translation, and self-driving cars are just a few examples.

Healthcare is considered the next domain to be revolutionized by Artificial Intelligence [2],[3],[4],[5]. In addition to many academic efforts, companies are also getting involved. IBM has developed *Watson* for several health applications, among which genomics, and there is a large number of start-ups addressing all possible aspects of the health continuum [6],[7].

**Artificial intelligence (AI)**

The term "artificial intelligence" is used to indicate development of algorithms that should execute tasks that are typically performed by human beings, and are therefore associated with intelligent behavior, making use of a variety of techniques, like deep learning, but also probabilistic methods like Bayesian modeling [8],[9];for definitions, see [2]. Colloquially, the term is applied to a machine that mimics "cognitive" functions, such as "learning" and "problem solving" [2],[4],[10],[11].

**Use of AI methods in healthcare**

It is clear that healthcare has numerous needs that could benefit from solutions developed with or embedding Artificial Intelligence [2],[4],[8]. In this brief article we focus on the contributions Artificial Intelligence can make to clinical healthcare, a domain that poses new and sometimes unique challenges to the application of Artificial Intelligence. In the next sections we discuss some important challenges and provide our recommendations on how to deal with them.

While radiology imaging was first in delivering digital data, digital pathology is a more recent revolutionary development [4],[12],[13],[14],[15]. In addition, for many years, hospitals have been digitizing their medical patient records [16]. Hence a large (and increasing) body of reasonably annotated clinical data has been collected, partially structured data (in machine readable formats, like from medical records), and partially unstructured data (in natural language). As in other industrial sectors, it is expected that this 'big data' movement can be leveraged to transform healthcare and drive unprecedented improvements in quality of patient diagnostics, care and clinical outcome. Expected results range from identification of individuals at high risk for a disease to improved diagnosis and matching of effective personalized treatment to the individual patient, as well as out-of-hospital monitoring of therapy response [17]. Although these opportunities and this potential are widely acknowledged, it is important to understand what can be delivered in practice with the current state-of-the-art AI technologies and which applications require further advances in AI to become feasible.

**AI methods: supervised or unsupervised learning; knowledge-based or data-driven?**

Multiple AI technologies are available to choose from [2],[9],[18]. Algorithmic learning-based AI can be performed in a supervised mode, which means that a 'ground truth' label is available for every data sample, which guides the AI effort and is based on domain knowledge. It will be obvious that the correctness of 'ground truth' labels is a prerequisite for good performance of an AI solution. The alternative, called, unsupervised, is when no ground truth is available and only similarities can be found with a yet undefined meaning [2],[18].

Machine learning traditionally involves a human to determine features of the data, using domain knowledge. In contrast, deep learning allows finding such features from the data 'by itself'. The features are subsequently used in various models. Some of those can be knowledge-based models in which new deep learning-defined features are integrated according to knowledge [19],[20]. The current interest in AI from industry comes from the recent breakthroughs in data-driven approaches, such as deep learning, and their applicability in industrial applications such as speech recognition, machine translation, computer vision, etc. Still, it is expected that combining data-driven and knowledge-based approaches will bring AI to the next level, much closer to human intelligence [19],[20].

**AI for well defined narrow tasks: radiology imaging and digital pathology**

One of the more studied and successfully executed AI opportunities is in imaging. For example AI technologies can be applied to problems such as distinguishing cell nuclei or certain cell types present in a tumor sample on a pathology slide, using slide images obtained with a digital pathology scanner [21],[22],[23]. Such images are made with consistent equipment and acquired in a controlled fashion, generating images consisting of uniform data, providing very good representation of the phenomena to be modelled. The problem domain is limited, i.e. in the training process the AI system gets as input raw images with associated labels for different cell types that are provided by a pathologist who in this way provides a 'ground truth', based on existing expert knowledge, for example on the different cell types or architecture present in the tissue slide. Deep learning has been applied to this problem and AI technologies already outperform manually crafted tissue analysis technologies [18],[24],[25][22]. It is expected that they will soon be on par or better than a human pathologist on certain well defined histology feature recognition and measurement tasks, though not yet for clinical interpretation.

**AI for wider purposes: coupling patient clinical data to clinical outcome**

On the other hand, research projects are ongoing using multimodal data, i.e. a combination of data sets of different nature, for example to enable prediction of prognosis of a patient or clinical outcome after a certain treatment. One may for example use medical imaging data combined with pathology and laboratory data and even lifestyle data to try to predict survival, x-day re-hospitalization-risks, etc. Such projects remain challenging and have typically been proven not to be very successful [5],[26]. IBM Watson claims to integrate all available patient data and disease information for improved diagnosis and therapy decision making. In 2013, the MD Anderson Cancer Center started using IBM Watson technology to increase effective treatment of cancer patients, however the project was stopped in 2017 because it "did not meet its goals" [27],[28]. In contrast, the concordance between single modality Watson for Genomics (WfG) and a genomics expert group for mutation analysis was reportedly quite good, between 77 and 97% depending on the type of mutations [7].

The most difficult challenge for AI in the coming years will be to move from (successful) narrow domains into wider-purpose, multi-modality, data systems. A promising approach here is not to find one methodology to address every problem but to separate the wider purpose AI goal into smaller goals. In this approach, sub-groups of the data may be processed separately with suitable AI methods to provide meaningful, clinically relevant, output. For example, for cardiac ultrasound images one could zoom in to develop an algorithm with deep learning to measure left ventricular volume, and for pathology slide images zoom in to develop an algorithm for recognition and quantification of a specific cell type, e.g. lymphocytes. To increase chances at success, it is important to determine a well-defined, focused and clinically relevant question for an AI project that can be adequately answered with the available data.

*Conclusion: Relevant and well-defined clinical question first.*

**The relationship between patient or sample numbers and data variables**

Many AI techniques, especially deep learning, rely on the availability of large data sets or "Big Data" [17]. Sometimes domain knowledge can help to create additional data derived from the data that are available. It is important to distinguish the type of data that is needed. In games, such as playing Chess and Go, it is easy to artificially synthesize additional data of the right type to increase the size of the dataset. With respect to medical and pathology imaging, large amounts of data are available on the premise that a relevant sample is defined as an image pixel, which is typically the case. Using this type of data, with relatively few images one can create millions of annotated samples with drawing tools. It is relatively easy to augment every sample with artificially generated variations (e.g. mirror copies, rotated versions, modified intensities, modified colors) without consequences for the annotation.

However, in clinical healthcare the type of data typically is a pathology or radiology image from a patient, associated with a clinical annotation like diagnosis or response to therapy. In this case the number of samples is generally equal to the number of patients. The annotation is often more difficult as it requires an expert physician to provide the 'ground truth'. When using multimodal data to find parameters that for example predict clinical outcome, despite all digital records and digital health devices, there is not 'enough' data. The number of patients for which the necessary multimodal data are available, is in general the limiting factor for using AI methods on such combined data sources to create a valid algorithm for risk prediction, diagnosis or a therapeutic decision. When the number of patients of a specific defined type is low, the often heard strategy is to extend a study to include more patients, even all patients worldwide, which requires addressing various legal and technical barriers. However, this is still likely to fail in reaching the required patient number, since with efforts to increase the number of patients for inclusion in data analysis, the amount of variation, including many unknown features/variables, per patient tends to grow as well, leading to uncontrolled data variation. This is caused by the large variation in human individuals: their DNA (just think of the 3 billion base pairs and the near infinite combinations of genomic variations), their lifestyle, (family) medical history, use of medication, etc. Moreover, patients are never treated in exactly the same manner in the various hospitals, bringing in many additional variables. It is a well-recognized issue in clinical trials run by pharma companies [26]. The challenge is to minimize such unwanted variables in the patient or sample set to analyze. Much of this uncontrolled variation is not recorded or at best only in a very noisy way. The number of unknown parameters that may have influenced the outcome, especially if its measurement lies many years after the diagnosis and treatment, is typically underestimated. A few examples of failure of AI methods caused by these issues are many genome-wide association studies aiming at identification of clinically useful genomic risk factors for complex diseases, and genomic studies aiming at identification of biomarkers for cancer diagnostics and treatment decisions [29].

Similar challenges are present in other domains, but there solutions can be invoked that are not possible in the healthcare domain. In natural language processing, Google Translate is a well-known example. When it started, translations were of very poor quality and heavily criticized, but Google decided to keep the service up and running, and on-line feedback was used to collect a large amount of translation data, enabling continuously improvement of the performance of the translation algorithm [30]. Summarizing, for applying AI to multimodal patient data, the number of patients from which the complete set of multimodal data is available is frequently too limited to address the 'curse of dimensionality'. In the scientific community, dimensionality reduction remains an active research area [31],[32],[33]. In clinical application areas for AI, it remains the main challenge to address. The first solution lies in reducing data modality and bringing the number of variables ($P$) on the right level in relation to the number of patients/samples ($N$) for which a 'ground truth' is available. The desired solution will reduce high dimension data to biologically sound knowledge-based features. Introducing knowledge-based computational approaches is expected to provide a way forward to reduce model freedom and handle high-dimensional data [34].

*Conclusion: Ratio between number of patients and their variables should fit the AI method.*

**Insufficient data quality and many subjective parameters**

For the patient data that is available, it turns out that this data is usually neither 100% complete nor 100% correct. For example, diagnoses are not always complete or correct, or they were not correctly entered in the digital domain. The main diagnosis is in general reasonably well documented, but side-diagnoses and complications, arising for example during hospital admission or in the home setting, treatment details, etc. are less accurately or not documented [5]. For many clinical variables like diagnoses, the ground truth comes from a physician's judgement and cannot be objectively measured or quantified. For example, it is documented that pathology diagnoses differ to a varying extent among pathologists that diagnose the same slide [15],[35],[36]. As a consequence, datasets may be incomplete and noisy, and presumed ground truths may not always be correct.

*Conclusion: Right data (representative and of a good quality) needs to be obtained.*

**Causal relations versus correlations: a role for Bayesian reasoning**

Any data-driven approach on data for which the ratio of number of patients (samples) to variables is too low can lead to multiple spurious correlations [37][37]. This means that the data suggest a correlation between two factors, but this is purely due to chance and there is no underlying explanation and causal relationship. In machine learning this can easily lead to overfitting and finding of irrelevant correlations [18],[38]. Also, for clinical implementation any interesting correlation (e.g. a feature or combination of features associated with increased disease risk) needs to be clinically validated at high cost, with very low success rates in view of the lack of causality. Therefore, turning an algorithm, based on correlations, into a successful proposition will in general be easier if causal relations underlie the found correlations. Knowledge-based reasoning techniques, like Bayesian network models, can reduce the number of spurious relations and overfitting problems by using existing knowledge on causal data relations to eliminate noisy data [11],[14]. As additional advantage Bayesian models can deal very well with uncertainty and missing variables, which is the rule rather than the exception in clinical data [11]. Not a coincidence, with respect to patient data interpretation they reason the way a medical doctor does [39].

*Conclusion: Relationship between data and ground truth should be as direct and causal as possible*

**Validation of AI-based solutions**

For many of the success stories of AI, a robust and reliable result is usually not necessary. For a free translation service, the consequence of a wrong decision is at most a dissatisfied customer. Improvements of those services could happen relatively fast because many of those AI applications are deployed in the field and iteratively improve their performance on the basis of new data learning from their mistakes. In sharp contrast to most of these consumer or lifestyle solutions based on AI, every clinical application, be it hardware or software, requires a thorough clinical validation in order to be adopted by the professional clinical community for use in patient care, like diagnostics or treatment decisions, and approved by regulatory authorities [26]. The requirements for clinical validation will be more stringent when errors or mistakes can have greater consequences. In a clinical trial it needs to be demonstrated how accurately the developed AI solution performs compared to the clinical standard, e.g. sensitivity and specificity of a diagnostic test. Still it is not completely clear whether good performance of an algorithm is acceptable if the solution is a 'black box' and not transparent and rationally explainable [2]. On top of that it is not obvious what proper validation of a continuous learning solution implies. An important issue is also that because of lack of transparency, deep learning-based 'black box' algorithms cannot be easily improved. This is in contrast to for example Bayesian models that are based on a transparent structure. Initial attemps to tackle this challenge are on the way [40].

Have AI-based solutions already been approved for clinical use? The earlier mentioned Watson Oncology system operates as a black box and its advice could not be clinically validated [28]. On the other hand, in 2017 it was claimed that the first deep-learning based algorithm, which identifies contours of cardiac ventricles from an MRI image to calculate ventricular volume, was validated and FDA approved for performing the calculation faster than a clinician [41]. Obviously, this system's scope is far more restricted than Watson's and the used unimodal imaging data are directly, and causally, related to the 'ground truth' provided during every image analysis by the clinician. Also, it can be considered a measurement algorithm, and does not include a clinical interpretation claim. Clinical validation and obtaining regulatory approval is much more difficult for those algorithms for which such an interpretation claim is added [4].

Several new solutions are ready for, or able to, perform continuous (or incremental) learning [42]. But within current regulations, an AI system for clinical applications should be 'frozen' and can therefore not learn online and 'immediately' apply its new knowledge. Rather, it needs to have an off-line validation of the obtained 'frozen' model on independent series of patient or sample datasets. Following a next continuous learning cycle, the validation process needs to be repeated again prior to renewed implementation of the model. Ideally, new clinically acceptable ways to shorten validation tracks for digital applications in a patient-safe manner should be found and it is expected that special procedures will be put in place to facilitate regulatory approval of updated algorithms. In line with this, the FDA is actively developing a strategy to deal with AI-based software solutions [43]. Maximal use of existing knowledge in transparent and causal model algorithms, like in Bayesian modeling, is expected to facilitate both clinical validation and obtaining regulatory approval, both for unimodal as well as for multimodal data.

*Conclusion: Regulatory ready; enabling validation*

**Methods to use**

Technology-wise numerous methods from the domain of AI have been explored for the development of clinical applications [8],[11],[18],[44]. Some have been more successful than others, mostly depending on application type. For automating pathology diagnosis using tissue slide images, deep learning has proven to be an appropriate technology. When dealing with more general, multimodal problems such as predicting clinical outcomes, patient assessments and risk predictions, currently other methods which often include domain knowledge are likely to be more appropriate choices. Probabilistic methods using knowledge representation are increasingly used and enable reduction of the number of influencing variables, determining 'sensible' features or latent variables. Probabilistic Bayesian modeling is well suited to deal with complex biological (e.g. "omics" data, like genomics, transcriptomics) and medical/clinical data, and is finding its way into diagnostic applications, as well as drug development [9],[45],[46],[47],[48],[49],[50]. However where knowledge is lacking, knowledge-agnostic AI approaches become valuable, and Bayesian reasoning networks are thought to have high potential for use in combination with deep learning, combining the best of two worlds in Bayesian deep learning [19],[20],[51].

*Conclusion: the Right AI method must be used for the problem.*

**Recommendations for use of AI-methods to develop clinical applications: The 6R's**

1. **Relevant and well-defined clinical question first.** Data analytics without domain knowledge can be applied in the healthcare domain, but at high risk of getting clinically irrelevant outcomes. For every new AI project, the clinical questions should be well defined and reviewed with clinical experts. The outcome of the analysis should also be reviewed for clinical and/or biological sense.
2. **Right data (representative and of a good quality)**. Carefully define the dataset that is needed to answer the clinical question. A clinical dataset with ground truth should be sufficiently clean and reliable. Be aware of hidden variation between samples that is not visible in the data set. The data set should be appropriate for the question at hand as well as representative for the population under study.
3. **Ratio between number of patients and their variables should fit the AI method.** To obtain useful results ensure working with adequately big data sets (numbers of patients/samples) for the AI method to be used, and reduce patient variables where possible. Use domain-knowledge to limit spurious correlations.
4. **Relationship between input variables and predicted output variable (as the dependent value) should be as direct and causal as possible.** The clinical question should as closely as possible relate the ground truth to the data. Hence, finding new pathology features that best distinguish between two different pathology diagnoses can be successful. Using lifestyle information to predict 10-year survival might not.
5. **Regulatory ready; enabling validation.** Upfront consider how a certain solution can be validated and pass regulatory requirements. Consider how using domain knowledge could speed up the validation process, for instance by breaking up the AI system into smaller AI systems. This effectively excludes systems that iteratively change by continuous learning.
6. **Right AI method.** Use the right method for the question at hand. Data-driven methods can be used if the data available allows it and knowledge-based methods can be applied if there is knowledge available but not enough data; a mixture of the two, combined in a wise manner, may be highly productive for development of clinically applicable healthcare solutions.

**Finally: privacy issues**

Driven by the Big Data analysis developments in healthcare, recently new privacy regulations were implemented in Europe (General Data Protection and Regulation, GDPR)[52]. To protect privacy, individuals control their own personal data, and explicit informed consent is required for access to the data and use in AI. This regulation is expected to make it more difficult to share patient data between multiple medical centers, and with companies involved in development of AI solutions.

**Key Takeaways**

While AI approaches are excellently suited to develop algorithms for analysis of unimodal imaging data, e.g. radiological or digital pathology images, for clinical (patient-related) applications major challenges lie in usually limited patient (sample) number ($N$) compared to the number of (multimodal) variables ($P$) (due to patient variation), inadequate 'ground truth' information, and a requirement for robust clinical validation prior to clinical implementation.

Artificial Intelligence solutions that combine domain knowledge with data-driven approaches are therefore preferable over solutions that use only domain knowledge or are fully data-driven.

We propose the following 6R model to keep in mind for AI projects:

1. Relevant and well-defined clinical question first
2. Right data (representative and of a good quality)
3. Ratio between number of patients and their variables should fit the AI method
4. Relationship between data and ground truth should be as direct and causal as possible
5. Regulatory ready; enabling validation

6. Right AI method

**Acknowledgement**

We wish to thank Rien van Leeuwen and Ruud Vlutters for their valuable contributions and Ludo Tolhuizen for thorough reading and providing valuable suggestions.